\documentclass{article}
\usepackage{amsmath}
\usepackage{graphicx}

\title{Some properties of a $\Delta-$string}
\author{V. Dzhunushaliev
\thanks{E-mail: dzhun@hotmail.kg}}

\begin{document}
\maketitle

\begin{center}
\textit{Dept. Phys. and Microel. Engineer., KRSU, Bishkek, \\
Kievskaya Str. 44, 720000, Kyrgyz Republic}
\end{center}

\begin{abstract}
The properties of 5D gravitational flux tubes are considered. 
With the cross section and 5th dimension in the Planck region such 
tubes can be considered as string-like objects, namely $\Delta-$strings. 
A model of attachment of $\Delta-$string to a spacetime is offered. 
It is shown that the attachment point is a model of an electric charge 
for an observer living in the spacetime. Magnetic charges are forbidden 
in this model. 
\end{abstract}

\section{Introduction}

Recently \cite{dzhun1} it was assumed that some kind of gravitational flux 
tube solutions in the 5D Kaluza-Klein gravity can be considered as string-like 
objects, namely $\Delta-$strings. The idea is that the gravitational flux 
tubes can be very thin (approximately $\approx l_{Pl}$) and arbitrary long. 
Evidently that such object looks like to a string attached to two Universes 
or to remote parts of a single Universe. The thickness of this object is so 
small that near to the point of attachment to an external Universe the handles 
of a spacetime foam will appear between this object and the Universe. 
This is like a delta of the river flowing into the sea. We call such objects as a 
$\Delta-$string. The $\Delta-$string has 
interesting properties : it can transfer electromagnetic waves and consequently 
the energy; the mouth is similar to a spread charge; it is a realization of 
Einstein-Wheeler's idea about ``mass without mass'' and ``charge without charge''; 
it has wormhole and string properties; in some sense it is some realization of 
Einstein's idea that all should be constructed from vacuum (in this case 
the $\Delta-$string is the vacuum solution of 5D gravitational equations). 
\par 
At first these gravitational flux tubes solutions were obtained as 4D 
Levi-Civita - Robertson - Bertotti solutions \cite{Levi-Civita} in 4D Einstein-Maxwell 
theory. It is an infinite flux tube filled with parallel electric $F_{01}$ and 
and magnetic $F_{23}$ fields 
\begin{eqnarray} 
ds^2 &=& a^2\left (\cosh^2 \zeta dt^2 - d\zeta^2 - d\theta ^2 -
\sin ^2 \theta d\varphi ^2\right ),
\label{intr-10} \\ 
F_{01} &=& \rho ^{1/2} \cos\alpha, \;\;\; \; \; \; \;
F_{23} = \rho ^{1/2}\sin\alpha, 
\label{intr-20} 
\end{eqnarray} 
where 
\begin{equation} 
G^{1/2} a \rho ^{1/2} = 1. 
\label{intr-30} 
\end{equation} 
$\alpha$ is an arbitrary constant angle; $a$ and $\rho$ are constants defined 
by Eq. (\ref{intr-30}); $a$ determines the cross section of the tube and $\rho$ 
the magnitude of the electric and magnetic fields; 
$G$ is Newton's constant ($c=1$, the speed of light); 
$F_{\mu\nu}$ is the electromagnetic field tensor. For $\cos\alpha = 1$ 
($\sin\alpha = 1$) one has a purely electric (or magnetic) field.  
\par 
In Ref. \cite{dzhsin1} was obtained similar 5D solutions. The properties 
of these solutions depend on the relation $\delta_{EH} = 1-E/H$ between electric 
$E$ and magnetic $H$ fields. The purpose of this paper is to investigate more careful 
the properties of the gravitational flux tubes with $0 < \delta_{EH} \ll 1$. 

\section{Numerical calculations}

\begin{figure}[h]
  \begin{minipage}[t]{.45\linewidth}
  \begin{center}
  \fbox{
  \includegraphics[height=5cm,width=5cm]{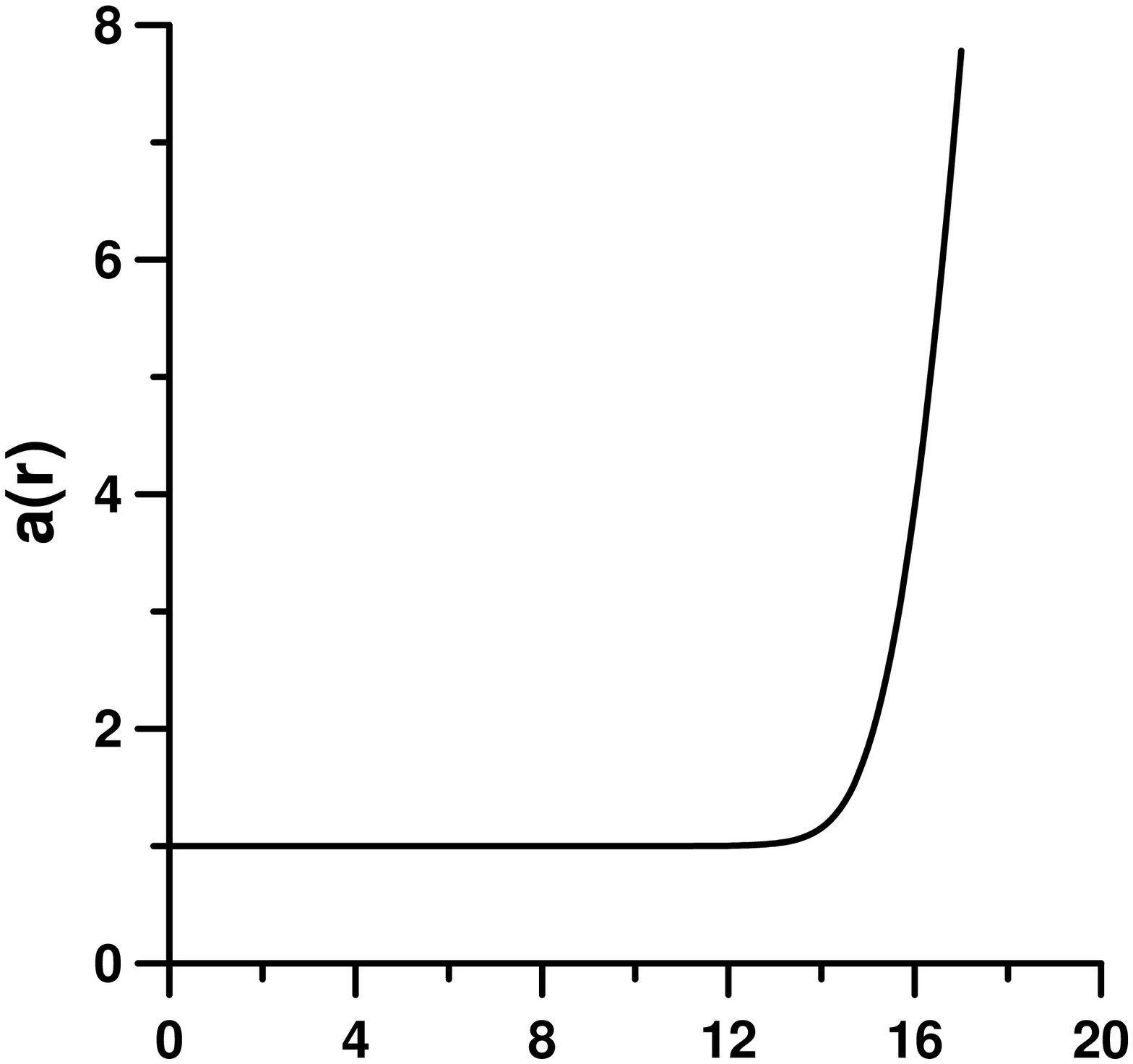}}
  \caption{The function $a(r)$. The parameter 
  $\delta_Q = 1 - \frac{Q}{Q_0} = 10^{-10}$.}
  \label{fig1}
  \end{center}
 \end{minipage}\hfill
 \begin{minipage}[t]{.45\linewidth}
 \begin{center}
  \fbox{
  \includegraphics[height=5cm,width=5cm]{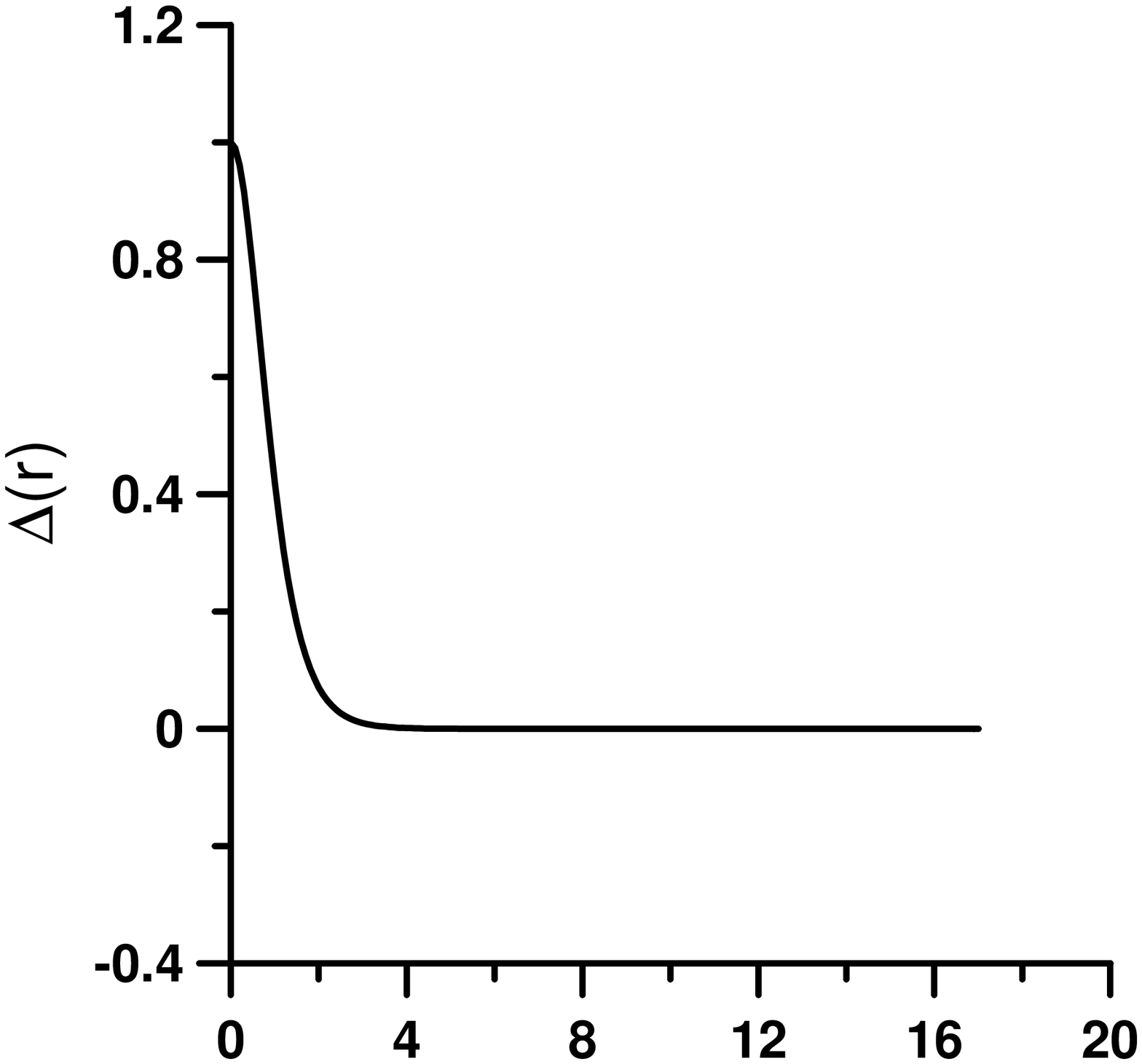}}
  \caption{The function $\Delta (r)$. The parameter 
  $\delta_Q = 1 - \frac{Q}{Q_0} = 10^{-10}$.}
  \label{fig2}
  \end{center}
 \end{minipage}\hfill
\end{figure}

\begin{figure}[h]
  \begin{minipage}[t]{.45\linewidth}
  \begin{center}
  \fbox{
  \includegraphics[height=5cm,width=5cm]{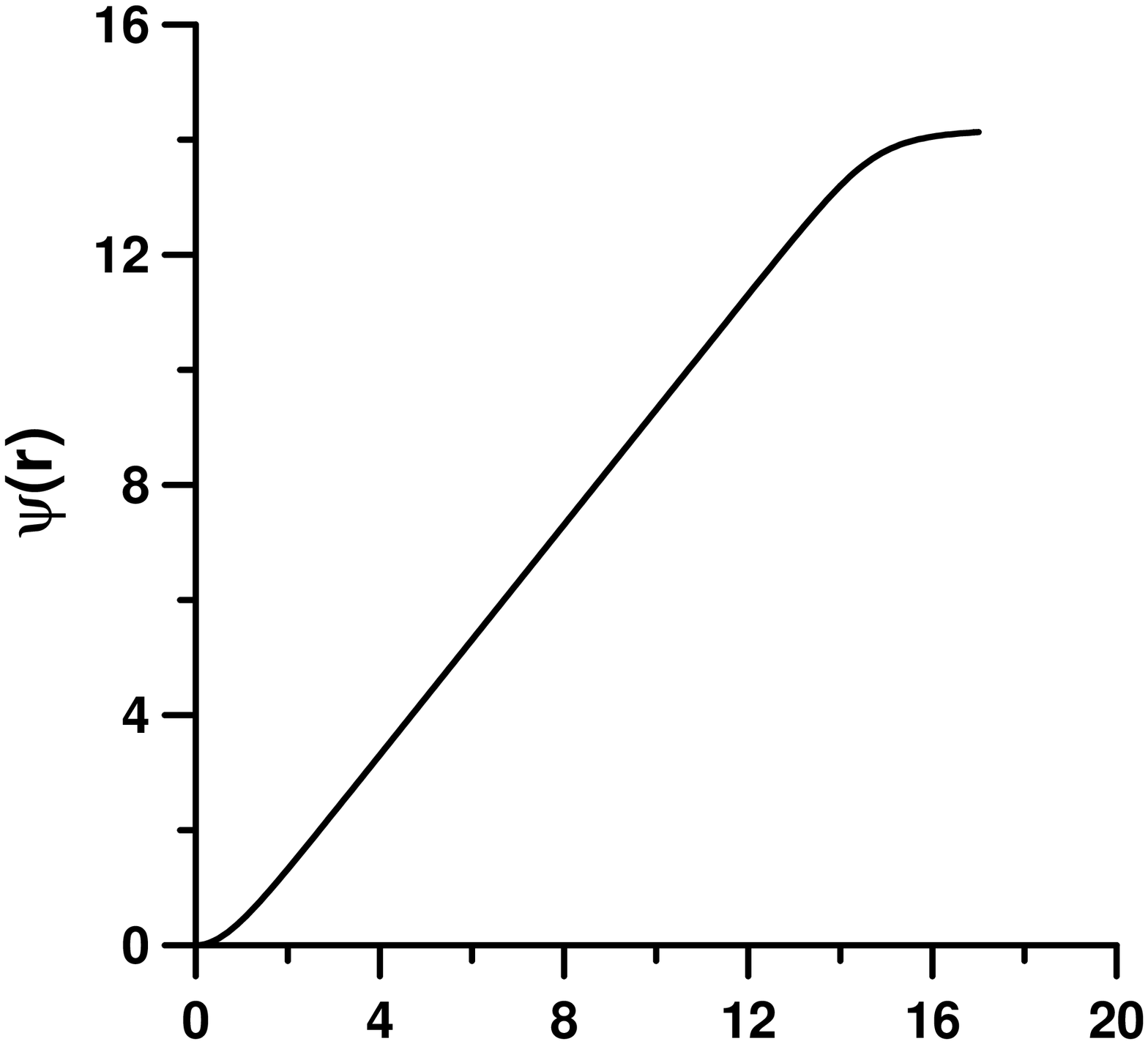}}
  \caption{The function $\psi (r)$. The parameter 
  $\delta_Q = 1 - \frac{Q}{Q_0} = 10^{-10}$.}
  \end{center}
  \label{fig3}
 \end{minipage}\hfill
  \begin{minipage}[t]{.45\linewidth}
  \begin{center}  
  \fbox{
  \includegraphics[height=5cm,width=5cm]{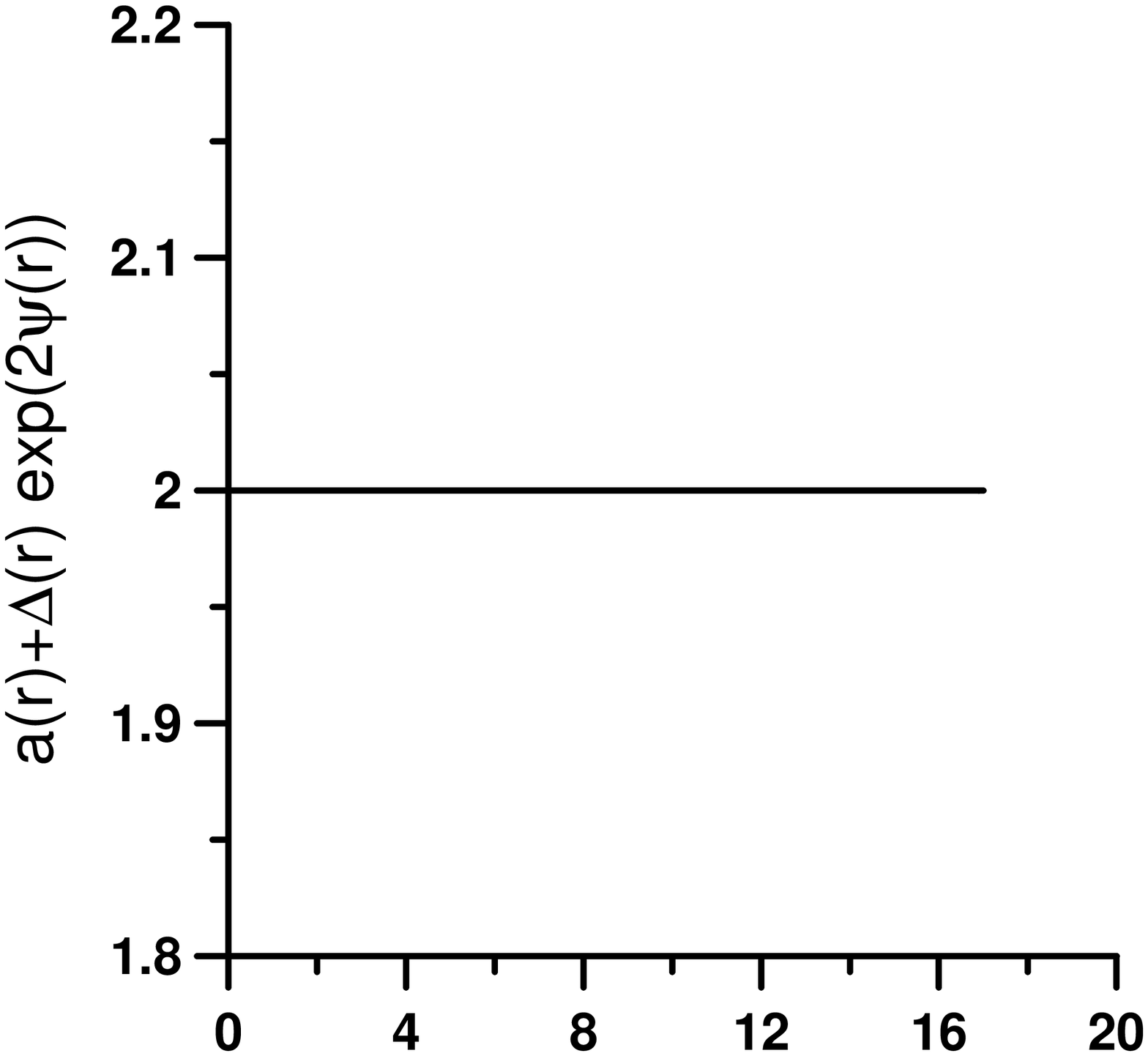}}
  \caption{The sum $a(r) + \Delta (r) e^{2\psi (r)}$. The parameter 
  $\delta_Q = 1 - \frac{Q}{Q_0} = 10^{-10}$.}
  \label{fig4}
  \end{center}
 \end{minipage}\hfill
\end{figure}

In this section we will investigate the solutions of 5D Einstein equations 
\begin{equation}
  R_{AB} - \frac{1}{2} \eta_{AB} R = 0
\label{sec2-10}
\end{equation}
here $A,B$ are 5-bein indices; $R_{AB}$ and $R$ are 5D Ricci tensor and the 
scalar curvature respectively; $\eta_{AB} = diag\{ 1,-1,-1,-1,-1 \}$. 
The investigated metric is 
\begin{equation}
%\begin{split}
  ds^2 = \frac{a(r)}{\Delta (r)} dt^2 - dr^2 - a(r) 
  \left(
  d\theta^2 + \sin^2 \theta d\varphi^2 
  \right) - 
%\\
  \frac{\Delta (r)}{a(r)} e^{2\psi (r)} 
  \left(
  d\chi + \omega (r) dt + Q \cos \theta d\varphi
  \right)^2 
%\end{split}  
\label{sec2-20}
\end{equation}
the functions $a(r), \Delta(r)$ and $\psi(r)$ are the even functions; 
$Q$ is the magnetic charge. The corresponding equations are 
\begin{eqnarray}
  R_{15} = \omega'' + \omega'
  \left(
  -\frac{a'}{a} + 2\frac{\Delta'}{\Delta} + 3 \psi '
  \right) &=& 0 ,
\label{sec2-30}\\
  R_{33} = \frac{a''}{a} +\frac{a'\psi'}{a} -\frac{2}{a} + 
  \frac{Q^2 \Delta e^{2\psi}}{a^3} &=& 0 ,
\label{sec2-40}\\
  R_{11} - R_{55} = \psi'' + {\psi'}^2 + \frac{a' \psi'}{a} - 
  \frac{Q^2 \Delta e^{2\psi}}{2a^3} &=& 0 ,
\label{sec2-50}\\
  2 R_{11} + R_{22} -R_{33} - R_{55} = 
  \frac{\Delta''}{\Delta} - \frac{\Delta' a'}{\Delta a} + 
  3\frac{\Delta' \psi'}{\Delta} + \frac{2}{a} - 6 \frac{a' \psi'}{a} &=& 0 ,
\label{sec2-60}\\
  R_{55} - R_{11} - R_{22} + 2R_{33} = 
  \frac{{\Delta'}^2}{\Delta^2} + \frac{4}{a} - 
  \frac{\Delta^2}{a^2} e^{2\psi}{\omega'}^2 - 
  \frac{Q^2 \Delta e^{2\psi}}{a^3} - 6\frac{a' \psi'}{a} - 
  2\frac{\Delta' a'}{\Delta a} + 2\frac{\Delta ' \psi '}{\Delta}&=& 0 .
\label{sec2-70} 
\end{eqnarray} 
The solution of Maxwell equation \eqref{sec2-30} is 
\begin{equation}
  \omega' = \frac{q a e^{-3\psi}}{\Delta^2} 
\label{sec2-75}
\end{equation}
here $q$ is the electric charge. It is interesting that equations 
\eqref{sec2-30}-\eqref{sec2-60} have not any contribution from the electric 
field \eqref{sec2-75}. Nevertheless the electric field has an influence on the 
solution due the equation for initial values \eqref{sec2-70} at the point $r=0$ 
\begin{equation}
  q^2 + Q^2 = 4 a(0)
\label{sec2-80}
\end{equation}
as $a'(0) = \Delta '(0) = \psi '(0) = 0$. The first analysis was done in 
Ref. \cite{dzhsin1} : the conclusion is that the solution depends on the 
magnitude of the magnetic charge $Q$. It is found that the solutions to 
the metric in Eq. (\ref{sec2-20}) evolve in the following way :
\begin{enumerate} 
\item 
$Q = 0$. The solution is \textit{a finite flux tube} 
located between two surfaces at $\pm r_H$ where $ds^2(\pm r_H) = 0$. 
The throat between the $\pm r_H$ surfaces is filled with electric flux.
\item 
$0 < Q < Q_0 = \sqrt{2 a(0)}$. The solution is again 
\textit{a finite flux tube}. The throat between the surfaces at 
$\pm r_H$ is filled with both electric and magnetic fields. 
\item 
$Q = Q_0$. In this case the solution is 
\textit{an infinite flux tube} filled with constant electrical and 
magnetic fields. The cross-sectional size of this solution is constant 
($ a= const.$). 
\item 
$Q_0 < Q < Q_{max} = 2 \sqrt{a(0)}$. In this case we have
\textit{a singular finite flux tube} located between two (+) and (-) 
electrical and magnetic charges located at $\pm r_{sing}$. At
$r = \pm r_{sing}$ this solution has real singularities which
we interpret as the locations of the charges. 
\item 
$Q = Q_{max}$. This solution is again
\textit{a singular finite flux tube} only with magnetic field filling the 
flux tube. In this solution the two opposite magnetic charges
are confined to a spacetime of fixed volume. 
\end{enumerate} 
The results of numerical calculations of equations 
\eqref{sec2-30}-\eqref{sec2-60} are presented on Fig. \ref{fig1}, 
\ref{fig2}, \ref{fig4} and \ref{fig4}. These numerical calculations allow 
us to suppose that there is the following relation between functions 
$a(r), \Delta(r)$ and $\psi(r)$ 
\begin{equation}
  a(r) + \Delta(r) e^{2\psi(r)} = 2 a(0) .
\label{sec2-90}
\end{equation}
In the next sections we will demonstrate the correctness of this relation 
by some approximate analytical calculations. 

\section{Approximate solution near to $Q=0$ solution}

In this section we will consider the metric in the form 
\begin{eqnarray}
  a(r) &=& a_0(r) + \delta a(r) ,
\label{sec3-10}\\
  \Delta(r) &=& \Delta_0(r) + \delta\Delta (r) ,
\label{sec3-20}\\
  \psi(r) &=& \psi_0(r) + \delta\psi(r) 
\label{sec3-30}
\end{eqnarray} 
and $Q$ is small parameter $Q^2/a(0) \ll 1$. 
\begin{eqnarray}
  a_0(r) &=& r^2 + r_0^2 ,
\label{sec3-40}\\
  \Delta_0(r) &=& r_0^2 - r^2 ,
\label{sec3-50}\\
  \psi_0(r) &=& 0 
\label{sec3-60}
\end{eqnarray} 
the functions $a_0(r), \Delta_0(r)$ and $\psi_0(r)$ are the solutions 
of Einstein equations with $Q=0$. The variation of equations 
\eqref{sec2-30}-\eqref{sec2-70} with respect to small perturbations 
$\delta a(r), \delta\Delta (r)$ and $\delta\psi(r)$ give us the following 
equations set 
\begin{eqnarray}
  \delta\psi + \frac{a'_0}{a_0}\delta\psi ' - 
  \frac{Q^2 \Delta_0}{2a_0^3} &=& 0 ,
\label{sec3-70}\\
  \delta a '' + a'_0 \delta\psi ' + \frac{Q^2 \Delta_0}{a_0^2} &=& 0 ,
\label{sec3-80}\\
  \delta\Delta '' - \frac{a'_0}{a_0}\delta\Delta'  - 
  \frac{\Delta'_0 \delta a'}{a_0} + 
  \frac{\Delta'_0 a'_0}{a_0^2} \delta a + 3 \Delta'_0 \delta\psi ' + 
  2\frac{\delta\Delta}{a_0} - 2\frac{\Delta_0}{a_0^2} \delta a &=& 0 .
\label{sec3-90}
\end{eqnarray} 
The solution of this system is 
\begin{eqnarray}
  \delta\psi(r) &=& -\frac{1}{4} \frac{Q^2}{r^2 + r_0^2} ,
\label{sec3-90}\\
  \delta a(r) &=& - \frac{Q^2}{2} \frac{r}{r_0} 
  \arctan \left( \frac{r}{r_0} \right) + \frac{Q^2}{4},
\label{sec3-100}\\
  \delta\Delta (r) &=& \frac{Q^2}{2} 
  \left[
  \frac{r}{r_0} \arctan \left( \frac{r}{r_0} \right) - 
  \frac{r^2 - r_0^2}{r^2 + r_0^2} + \frac{1}{2} 
  \right] .
\label{sec3-110}\\
\end{eqnarray} 
Substituting into equation \eqref{sec2-90} gives us 
\begin{equation}
  a(r) + \Delta (r) e^{2\psi(r)} = 2r_0^2 + \frac{Q^2}{2}
\label{sec3-120}
\end{equation}
but 
\begin{equation}
  a(r) \approx r_0^2 + r^2 - \frac{Q^2}{2} \frac{r}{r_0} 
  \arctan \left( \frac{r}{r_0} \right) + \frac{Q^2}{4}
\label{sec3-130}
\end{equation}
and 
\begin{equation}
  a(0) \approx r_0^2 + \frac{Q^2}{4}
\label{sec3-140}
\end{equation}
therefore 
\begin{equation}
  a(r) + \Delta (r) e^{2\psi(r)} = 2a(0)
\label{sec3-120}
\end{equation}
with an accuracy of $Q^2$. 

\section{Expansion into a series}

In this section we will derive the solution by expansion in terms of $r$. 
Using the MAPLE package we have obtained the solution with accuracy of 
$r^{12}$
\begin{eqnarray}
  \psi(x) &=& \frac{1}{4}{\tilde{Q}}^{2}{x}^{2} + 
  \left (
  {\frac {5}{48}} {\tilde{Q}}^{4} - \frac{1}{4} {\tilde{Q}}^{2}
  \right ){x}^{4} + 
  \left (
  -{\frac {7}{30}}{\tilde{Q}}^{4} + {\frac {41}{720}}{\tilde{Q}}^{6} + 
  \frac{1}{4} {\tilde{Q}}^{2}
  \right ){x}^{6}
\nonumber\\
  && + \left (-\frac{1}{4}{\tilde{Q}}^{2}+{\frac {213}{560}} {\tilde{Q}}^{4}-{
  \frac {111}{560}} {\tilde{Q}}^{6}+{\frac {281}{8064}} {\tilde{Q}}^{8}
  \right ){x}^{8}+
\nonumber\\
  && \left (-{\frac {341}{630}} {\tilde{Q}}^{4} + 
  {\frac {3749}{8400}} {\tilde{Q}}^{6} - {\frac {1553}{9450}} {\tilde{Q}}^{8}+
  {\frac {5147}{226800}} {\tilde{Q}}^{10} + \frac{1}{4}{\tilde{Q}}^{2}\right ){x}^
  {10}+O\left ({x}^{12}\right ),
\label{sec4-10}\\
  \Delta(x) &=& 
  1-{x}^{2}+
  \left (\frac{5}{6} {\tilde{Q}}^{2}-\frac{1}{4} {\tilde{Q}}^{4}
  \right ){x}^{4}+
  \left (
  -{\frac {9}{10}} {\tilde{Q}}^{2}+{\frac {47}{90}} {\tilde{Q}}^{4}-\frac{1}{12} {\tilde{Q}}^{6}
  \right ){x}^{6} + 
\nonumber\\ 
  && \left (-\frac{6}{7} {\tilde{Q}}^{4}+
  {\frac {373}{1260}} {\tilde{Q}}^{6}-{\frac {3}{80}} {\tilde{Q}}^{8}+{
  \frac {13}{14}} {\tilde{Q}}^{2}\right ){x}^{8} + 
\nonumber\\ 
  && \left ({\frac {218}{175}} {\tilde{Q}}^{4}-
  {\frac {2893}{4200}} {\tilde{Q}}^{6}+{\frac {5219}{
  28350}} {\tilde{Q}}^{8}-{\frac {17}{18}} {\tilde{Q}}^{2}-{\frac {37}{1890}} {\tilde{Q}}^{10}
  \right ){x}^{10} + O\left ({x}^{12}\right ),
\label{sec4-30}\\ 
  a(x) &=& 
  1+\left (1-\frac{1}{2} {\tilde{Q}}^{2}\right )
  \left[
  {x}^{2} + \frac{1}{6} {\tilde{Q}}^{2}{x}^{4} + 
  \left (-\frac{1}{10} {\tilde{Q}}^{2}+{\frac {11}{180}} {\tilde{Q}}^{4}\right ){x}^{6} + 
  \right .
\nonumber \\
  &&\left .
  \left (-{\frac {13}{140}} {\tilde{Q}}^{4}+{\frac {73}
  {2520}} {\tilde{Q}}^{6}+\frac{1}{14} {\tilde{Q}}^{2}\right ){x}^{8} + 
  \right .
\nonumber \\
  &&\left .  
  \left ({\frac {239}{2100}} {\tilde{Q}}^{4}+{\frac {887}{56700}} 
  {\tilde{Q}}^{8}-\frac{1}{18} {\tilde{Q}}^{2}-{\frac {13}{175}} {\tilde{Q}}^{6}
  \right ){x}^{10}
  \right] + O\left ({x}^{12}\right ) 
\label{sec4-40}
\end{eqnarray}  
here we have introduced the following dimensionless parameters : $x = r/\sqrt{a(0)}$ 
and $\tilde{Q} = Q/\sqrt{a(0)}$. The substitution in eq. \eqref{sec2-90} shows that 
it is valid. 
\par 
One of the most important question here is : how long is the $\Delta-$string ? 
Let us determine the length of the $\Delta-$string as $2 r_H$ where $r_H$ is 
the place where $ds^2 (\pm r_H) = 0$ (at these points $\Delta (\pm r_H) = 0$ and 
$a(\pm r_H) = 2 a(0)$). This is very complicated problem because we have 
not the analytical solution and the numerical calculations indicate that 
$r_H$ depends very weakly from $\delta_Q^{-1} = (1 - Q/Q_0)^{-1}$ parameter : 
the big magnitude of this parameter leads to the relative small magnitude of 
$r_H$. The expansion \eqref{sec4-40} allows us to assume that 
\begin{equation}
  a(r) = a(0) +  a_1(r, Q) \, \delta
\label{sec4-50}
\end{equation}
here $\delta = 1 - Q^2/Q_0^2$; $Q_0^2 = 2 a(0)$. If 
$Q \approx Q_0 = \sqrt{2a(0)}$ then 
\begin{equation}
  a(r) \approx a(0) +  a_1(r, Q_0) \, \delta .
\label{sec4-60}
\end{equation}
Using the MAPLE package we have obtained $a_1(r)$ with accuracy of $r^{20}$ 
\begin{equation}
\begin{split}
  a_1(x) = {x}^{2} + \frac{1}{3} {x}^{4} + 
  {\frac {2}{45}} {x}^{6} + 
  {\frac {1}{315}} {x}^{8} + 
  {\frac {2}{14175}} {x}^{10} +  
\\
  {\frac {2}{467775}} {x}^{12} + 
  {\frac {4}{42567525}} {x}^{14} + 
  {\frac {1}{638512875}} {x}^{16} +
  {\frac {2}{97692469875}} {x}^{18} + 
  O\left ({x}^{20}\right ) . 
\label{sec4-70}
\end{split}
\end{equation}
It is easy to see that this series coincides with $\cosh^2(x)-1$. 
Therefore we can assume that in the first rough approximation 
\begin{equation}
  a(r) \approx a(0) 
  \left \{ 1 + 
  \left(
  1 - \frac{Q^2}{Q_0^2}
  \right)
  \left[
  \cosh^2 \left( \frac{r}{\sqrt{a(0)}}\right) - 1 
  \right] 
  \right \}.
\label{sec4-80}
\end{equation}
It allows us to estimate the length $L = 2 r_H$ of the $\Delta-$string from 
equation \eqref{sec2-90} as 
\begin{equation}
  a(r_H) = a(0) 
  \left \{
  1 + \delta \left[ \cosh^2 \left( \frac{r_H}{\sqrt{a(0)}}\right)  - 1 \right]
  \right \} = 2a(0) .
\label{sec4-90}
\end{equation}
For $r_H \gg \sqrt{a(0)}$ we have 
\begin{equation}
  L \approx \sqrt{a(0)} \ln \frac{1}{\delta} .
\label{sec4-100}
\end{equation}
One can compare this result with the numerical calculations : 
$\delta = 10^{-10}$, $r_H/\sqrt{a(0)} \approx 15$, 
$\frac{1}{2}\ln 1/\delta \approx 15$; 
$\delta = 10^{-9}$, $r_H/\sqrt{a(0)} \approx 11$, 
$\frac{1}{2}\ln 1/\delta \approx 10$; 
$\delta = 10^{-8}$, $r_H/\sqrt{a(0)} \approx 10$, 
$\frac{1}{2}\ln 1/\delta \approx 9$. 
We see that $r_H/\sqrt{a(0)}$ and $\ln 1/\delta$ have the same order. 
This evaluation should be checked up (in future investigations) more 
carefully as the convergence radius of our expansion 
\eqref{sec4-10}-\eqref{sec4-40} is unknown. 

\section{The model of the $\Delta-$string}

\begin{figure}[h]
  \begin{center}
  \fbox{
  \includegraphics[height=5cm,width=7cm]{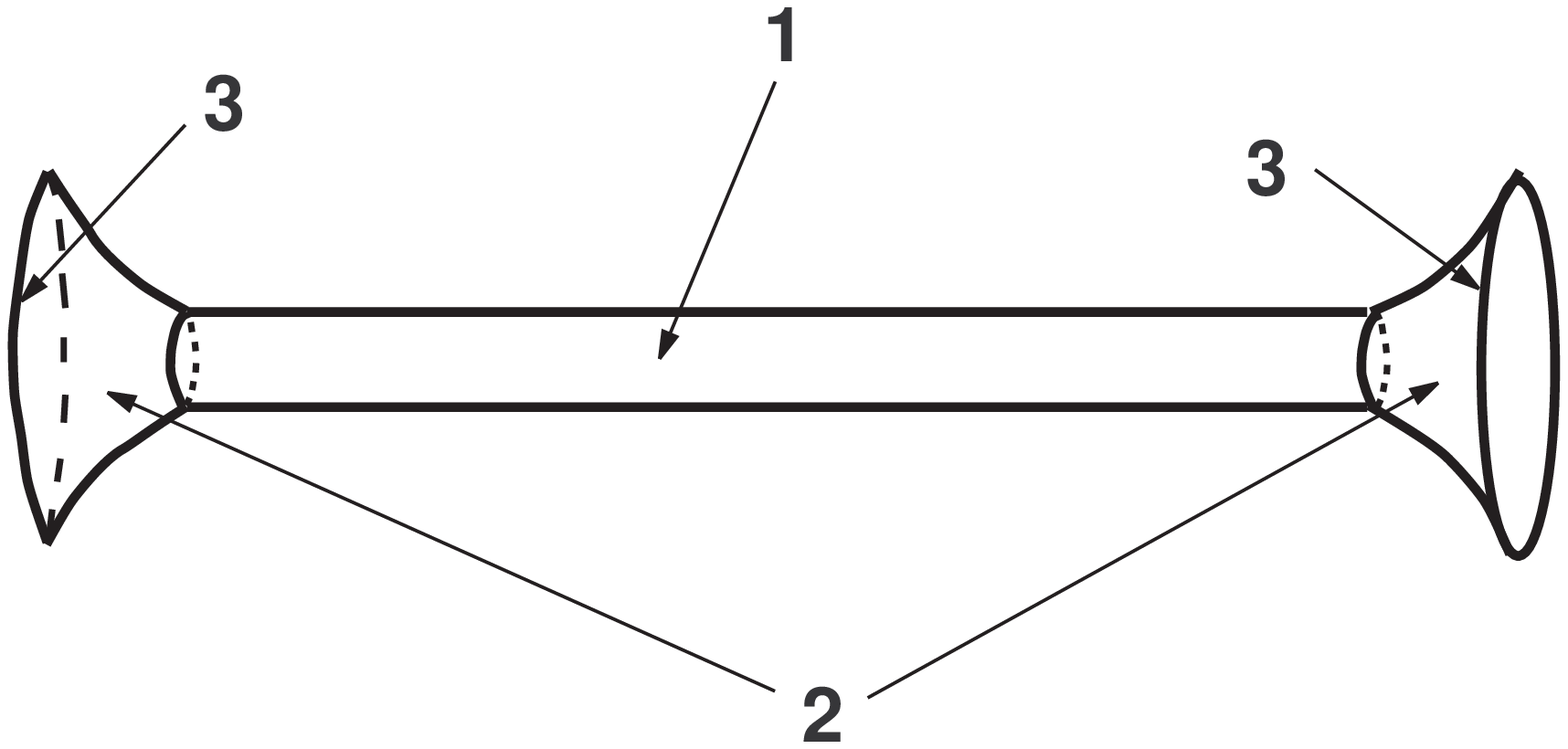}}
  \caption{The approximate model of the $\Delta-$string. 
  \textbf{1} is the long tube which is the part of the $Q = Q_0$ 
  solution. \textbf{2} are two cones which are the $Q=0$ solutions, 
  \textbf{3} are two hypersurfaces where $ds^2=0$.}
  \label{fig5}
  \end{center}
\end{figure}

The numerical calculations presented on Fig's \eqref{fig1}-\eqref{fig4} 
show that the $\Delta-$string approximately can be presented as a finite tube 
with the constant cross section and a big length joint at the ends with two short 
tubes which have variable cross section, see Fig. \eqref{fig5}. 
\par 
For the model of the central tube we take the part of the infinite flux tube 
solution with $Q = Q_0$ \cite{dzhsin1} 
\begin{eqnarray} 
a_\infty = \frac{Q_0^2}{2} = const, 
\label{sec5-10}\\
e^{\psi} = \frac{a_\infty}{\Delta} = \cosh\frac{r}{\sqrt{a_\infty}},
\label{sec5-20}\\
\omega = \sqrt{2}\sinh\frac{r}{\sqrt{a_\infty}} 
\label{sec5-30}
\end{eqnarray} 
here we have parallel electric $E$ and magnetic $H$ fields with equal electric 
$q$ and magnetic $Q$ charges. 
\par 
For a model of two peripheral ends (cones) we take the solution with $Q=0$ 
(equations \eqref{sec3-40}-\eqref{sec3-60}). Here we have only the electric 
field $E$ - 
\eqref{sec2-75}. At the ends of the $\Delta-$string ($r = \pm r_H$) 
the function 
$\Delta(\pm r_H) = 0$ and the term $Q^2\Delta e^{2\psi}/a^3$ in equations 
\eqref{sec2-40}-\eqref{sec2-70} is zero that allows us to set this term as zero 
at the peripheral cones. 
\par 
We have to join the components of metric on the $r=r_1=r_H-\sqrt{a(0)}$ 
(here $a(0) = a_\infty$). 
\begin{equation}
  \cosh^2 \left( \frac{r_1}{\sqrt{a(0)}} \right) = 
  \frac{\tilde{a}(r_1)}{\tilde{\Delta}(r_1)} = 
  \tilde{g}_{tt}(r_1) = 
  \tilde{\tilde{g}}_{tt}(r_1) = 
  \frac{\tilde{\tilde{a}}(r_1)}{\tilde{\tilde{\Delta}}(r_1)}
\label{sec5-40}
\end{equation}
here $\tilde{ }$ and $\tilde{\tilde{ }}$ mean that the corresponding quantities 
are the metric components for $Q=Q_0$ and $Q=0$ solutions respectively. According 
to equations \eqref{sec3-40}-\eqref{sec3-60} 
\begin{eqnarray}
  \tilde{\tilde{a}}(r) &=& a(0) + \left( r - r_1 \right)^2 , 
\label{sec5-50}\\
  \tilde{\tilde{\Delta}}(r) &=& \alpha 
  \left[
  a(0) - \left( r - r_1 \right)^2 
  \right]
\label{sec5-60}\\ 
  \tilde{\tilde{a}}(r_1) &=& \tilde{a}(r_1) = a(0) 
\label{sec5-65}
\end{eqnarray}  
here $r_0^2$ from equation \eqref{sec3-40}-\eqref{sec3-60} is replaced with 
$\tilde{\tilde{a}}(0)$ and some coefficient $\alpha$ is introduced as equations 
\eqref{sec2-40}-\eqref{sec2-70} with $Q=0$ have the terms like 
$\Delta'' /\Delta$ and $\Delta'/\Delta$ only. It gives us 
\begin{equation}
  \alpha = \frac{1}{\cosh^2\left( \frac{r_1}{\sqrt{a(0)}} \right)} 
  \approx 4e^{\frac{-2 r_1}{\sqrt{a(0)}}}
\label{sec5-70}
\end{equation}
since for the long $\Delta-$string $r_1/\sqrt{a(0)} \gg 1$. The next joining 
is for 
\begin{equation}
  1 = \frac{\tilde{\Delta}(r_1) e^{2 \tilde{\psi}(r_1)}}{\tilde{a}(r_1)} = 
  \tilde{g}_{55}(r_1) = \tilde{\tilde{g}}_{55}(r_1) = 
  \frac{\tilde{\tilde{\Delta}}(r_1) e^{2 \tilde{\tilde{\psi}}(r_1)}}
  {\tilde{\tilde{a}}(r_1)}
\label{sec5-80}
\end{equation}
here the constant term $2 \tilde{\tilde{\psi}}_1$ is added to solution 
\eqref{sec3-60} $2\psi_0=0$ since again equations \eqref{sec2-40}-\eqref{sec2-70} 
have $\psi''$ and $\psi'$ terms only. Consequently 
\begin{equation}
  e^{2\tilde{\tilde{\psi}}_1} = \frac{1}{\alpha} \approx 
  \frac{e^{2 r_1/\sqrt{a(0)}}}{4} \quad 
  \text{or} \quad 
  \psi_1 \approx \frac{r_1}{\sqrt{a(0)}} - \ln 4 .
\label{sec5-90}
\end{equation}
The last component of the metric is 
\begin{equation}
  a(0) = a_\infty = \tilde{g}_{\theta\theta}(r_1) = 
  \tilde{\tilde{g}}_{\theta\theta}(r_1) = \tilde{\tilde{a}}(r_1) .
\label{sec5-100}
\end{equation}
For the electric field we have 
\begin{eqnarray}
  \tilde{E}(r_1) &=& \frac{q e^{\tilde{\psi}(r_1)}}{a(0)} = 
  \frac{q}{a(0)} \cosh \left( \frac{r_1}{\sqrt{a(0)}} \right) = 
  \frac{q}{a(0)} \frac{1}{\sqrt{\alpha}}
\label{sec5-110}\\
  \tilde{\tilde{E}}(r_1) &=& q \frac{\tilde{\tilde{a}}(r_1)}
  {\tilde{\tilde{\Delta}}^2(r_1)} 
  e^{-3 \tilde{\tilde{\psi}}(r_1)} = 
  \frac{q}{a(0)} \frac{1}{\sqrt{\alpha}}
\label{sec5-120}
\end{eqnarray}  
and consequently 
\begin{equation}
  \tilde{E}(r_1) = \tilde{\tilde{E}}(r_1) .
\label{sec5-130}
\end{equation}
Thus our final result for this section is that at the first rough approximation 
the $\Delta-$string looks like to a tube with constant cross section and two 
cones attached to its ends. 

\section{$\Delta-$string as a model of electric charge}

In this section we want to present a model of attachment of the 
$\Delta-$string to a spacetime. 
\par 
Maybe the most important question for the $\Delta-$string is : what will see 
an external observer living in the Universe to which the $\Delta-$string 
is attached. Whether he will see a dyon with electric and magnetic fields 
or an electric charge ? This question is not very simple as it is not very 
clear what is it the electric and magnetic fields on the $\Delta-$string. 
Are they the tensor $F_{\mu\nu}$ and fields $E_i$ and $H_i$ 
\begin{eqnarray}
  F_{\mu\nu} &=& \partial_\mu A_\nu - \partial_\nu A_\mu ,
  \quad \mu , \nu = 0,1,2,3 ,
\label{sec6-10}\\
  E_i &=& F_{0i}, \quad i,j,k = 1,2,3 ,
\label{sec6-20}\\
  H_i &=& e_{ijk} F^{jk} 
\label{sec6-30}  
\end{eqnarray}
or something another that is like to an electric displacement 
$D_i = \varepsilon E_i$ and $H_i = \mu B_i$ ? The answer on this question 
depends on the way how we will continue the solution behind the hypersurface 
$r=\pm r_H$ (where $ds^2=0$). We have two possibilities : the first 
way is the simple continuation our solution to $|r|>r_H$, the second way is a 
string approach - attachment the $\Delta-$string to a spacetime. In the 
first case $\Delta(r) > 0$ by $|r|<r_H$ and $\Delta(r) < 0$ by $|r|>r_H$ 
it means that by $|r|>r_H$ the time and 5th coordinate $\chi$ becomes 
respectively space-like and time-like dimensions. It is not so good for us. 
The second approach is like to a string attached to a D-brane. We believe that 
physically the second way is more interesting. 
\par 
Let us consider Maxwell equations in the 5D Kaluza-Klein theory 
\begin{equation}
  R_{5\mu} = \frac{1 }{\sqrt{-g}} \partial_\nu 
  \left(
  \sqrt{-g} g_{55} F^{\mu\nu}
  \right) = 0
\label{sec6-40}
\end{equation}
here $g = \det{g_{AB} = a^4 \sin^2 \theta e^{2\psi}}$. These 5D Maxwell 
equations are similar to Maxwell equations in the continuous media where 
the factor $\sqrt{-g} g_{55}$ is like to a permittivity for 
$R_{50}$ equation and a permeability for $R_{5i}$ equations. It 
allows us offer the following conditions for matching the electric and magnetic 
fields at the attachment point of $\Delta-$ string to the spacetime 
\begin{eqnarray}
  \sqrt{-g_\Delta} {g_{\Delta}}_{55} F^{0i}_\Delta &=& \sqrt{-g} F^{0i} 
\label{sec6-50}\\
  \sqrt{-g_\Delta} {g_{\Delta}}_{55} F^{ij}_\Delta &=& \sqrt{-g} F^{ij} 
\label{sec6-60}
\end{eqnarray} 
here the subscript $\Delta$ means that this quantity is given on the 
$\Delta-$string and 
on l.h.s there are the quantities belonging to the $\Delta-$string 
and on the r.h.s. the corresponding quantities belong to the spacetime 
to which we want to attach the $\Delta-$string. As an example one can 
join the $\Delta-$string to the Reissner-Nordstr\"om solution with the 
metric 
\begin{equation}
  ds^2 = e^{2\nu} dt^2 - e^{-2\nu} dr^2 - r^2 
  \left( d\theta^2 + \sin^2\theta d\varphi^2 \right).
\label{sec6-70}
\end{equation}
Then on the $\Delta-$string 
\begin{equation}
  \sqrt{-g_\Delta} {g_{\Delta}}_{55} F^{tr}_\Delta = 
  - \frac{\Delta^2 e^{3\psi}}{a} \omega ' = 
  -a \frac{q}{a}
\label{sec6-80}
\end{equation}
For the Reissner-Nordstr\"om solution 
\begin{equation}
  \sqrt{-g} F^{tr} = - r^2 E_{r(RN)}.
\label{sec6-90}
\end{equation}
Therefore after joining on the event horizon $r=r_H$ ($a=a_H$) we have 
\begin{equation}
  E_{r(RN)} = \frac{q}{a_H}
\label{sec6-100}
\end{equation}
here we took into account that $a_H=r^2_H$. 
\par 
For equation \eqref{sec6-60} we have 
\begin{equation}
  \sqrt{-g_\Delta} {g_{\Delta}}_{55} F^{\theta \varphi}_\Delta = 
  \frac{Q \Delta e^{3\psi}}{a^2}.
\label{sec6-110}
\end{equation}
On the surface of joining $\Delta(r_H) = 0$ consequently we have very 
unexpected result : the continuation of $F^{\theta\varphi}$ 
(and magnetic field $H_r$) from the $\Delta-$string to the spacetime 
(D-brane) gives us zero magnetic field $H_r$. 
\par 
Finally our result for this section is that the $\Delta-$string can be 
attached to a spacetime by such a way that it looks like to an electric 
charge but not magnetic one. Such approach to the geometrical interpretation 
of electric charges suddenly explain why we do not observe magnetic charges 
in the world. 

\section{Discussion and conclusions}

In this paper we have considered some interesting properties of the 
$\Delta-$string. We have shown that there is the relation \eqref{sec2-90} 
between the metric components. Physically this relation shows that the 
cross section of $\Delta-$string is the same order on the interval 
$(-r_H , +r_H)$. In fact we can choose the cross section as $\approx l_{Pl}$ 
that guarantees us that the cross section of $\Delta-$string is in Planck 
region everywhere between points $\pm r_H$ where the $\Delta-$string can be 
attached to D-branes. 
\par 
It is shown that in some rough approximation the length of $\Delta-$string 
depends logarithmically from a small number which is connected with 
the difference between $1 - Q/Q_0$. 
\par 
The calculations presented here give us an additional certainty that the 
string can have an inner structure and in gravity such string is the 
$\Delta-$string. This approach is similar to Einstein point of view that 
the electron has an inner structure which can be described inside of a 
gravitational theory as a vacuum topologically non-trivial solution. 
\par 
Another interesting peculiarity is that the gravitational flux tube 
solutions appear in a relatively simple matter in the classical 5D 
gravity. This is in contrast with quantum chromodynamic where receiving 
a flux tube stretched between quark and antiquark is an unresolved 
problem. In both cases strings are some approximation for the tubes. 
It is possible that it is a manifestation of some nontrivial mapping 
between a classical gravity and a quantum field theory. 
\par 
It is necessary to note that similar idea was presented in Ref. 
\cite{guendelman} : the matching of two remote regions was done using 4D 
infinite flux tube which is the Levi-Civita - Bertotti - Robinson solution 
filled with the electric and magnetic fields. 

\section{Acknowledgment}
I am very grateful to the ISTC grant KR-677 for the financial support.

\end{document}